\documentclass[conference,retainorgcmds]{IEEEtran}
\IEEEoverridecommandlockouts
\usepackage{cite}
\usepackage{amsmath,amssymb,amsfonts}
\usepackage{algorithm,algorithmic}
\usepackage{graphicx}
\usepackage{textcomp}
\usepackage{url}

\usepackage{tablefootnote}
\usepackage{multirow}
\usepackage{multicol}
\usepackage{soul,xcolor}
\usepackage{caption}
\usepackage{subcaption}
\usepackage[normalem]{ulem}

\graphicspath{{figures/}} 

\begin{document}

\title{Reinforcement Learning Based Resource Allocation for Network Slices in O-RAN Midhaul}
\author{\IEEEauthorblockN{Nien Fang Cheng, \IEEEmembership{Student Member, IEEE}, Turgay Pamuklu, \IEEEmembership{Member, IEEE},\\ Melike Erol-Kantarci, \IEEEmembership{Senior Member, IEEE}}

\IEEEauthorblockA{\textit{School of Electrical Engineering and Computer Science,}
\textit{University of Ottawa}, Ottawa, Canada}

\IEEEauthorblockA{
Emails:\{nchen048, turgay.pamuklu, melike.erolkantarci\}@uottawa.ca}
}
\maketitle
\makeatletter
\def\ps@IEEEtitlepagestyle{%
  \def\@oddfoot{\mycopyrightnotice}%
  \def\@oddhead{\hbox{}\@IEEEheaderstyle\leftmark\hfil\thepage}\relax
  \def\@evenhead{\@IEEEheaderstyle\thepage\hfil\leftmark\hbox{}}\relax
  \def\@evenfoot{}%
}
\def\mycopyrightnotice{%
  \begin{minipage}{\textwidth}
  \centering \scriptsize
Accepted Paper. IEEE policy provides that authors are free to follow funder public access mandates to post accepted articles in repositories. When posting in a repository, the IEEE embargo period is 24 months. However, IEEE recognizes that posting requirements and embargo periods vary by funder. IEEE authors may comply with requirements to deposit their accepted manuscripts in a repository per funder requirements where the embargo is less than 24 months.
  \end{minipage}
}
\makeatother

\begin{abstract}
Network slicing envisions the 5th generation (5G) mobile network resource allocation to be based on different requirements for different services, such as  Ultra-Reliable Low Latency Communication (URLLC) and Enhanced Mobile Broadband (eMBB). Open Radio Access Network (O-RAN), proposes an open and disaggregated concept of RAN by modulizing the functionalities into independent components. Network slicing for O-RAN can significantly improve performance.  Therefore, an advanced resource allocation solution for network slicing in O-RAN is proposed in this study by applying Reinforcement Learning (RL). This research demonstrates an RL compatible simplified edge network simulator with three components, user equipment(UE), Edge O-Cloud, and Regional O-Cloud. This simulator is later used to discover how to improve throughput for targeted network slice(s) by dynamically allocating unused bandwidth from other slices. Increasing the throughput for certain network slicing can also benefit the end users with a higher average data rate, peak rate, or shorter transmission time. The results show that the RL model can provide eMBB traffic with a high peak rate and shorter transmission time for URLLC compared to balanced and eMBB focus baselines.
\end{abstract}

\section{Introduction}
The new era of entering the 5th generation (5G) mobile networking started with 3GPP Release 15. Later on 5G New Radio (5G NR) technology was defined in more detail in Release 16 with more advanced capabilities \cite{3GPPREl16}. This new definition of Radio Access Network (RAN) has introduced possibilities of redefining the resource allocation for traffic on the edge network.

In 5G NR, network slicing helps fulfilling different Key Performance Indicators (KPI) required by various service types. There are three new characteristic services in 5G, enhanced Mobile Broadband (eMBB), Ultra-Reliable and Low Latency (URLLC), and massive Machine Type Communications (mMTC). They represent different needs from the user equipment (UE). eMBB is looking for a high peak rate, while URLLC demands low delays and mMTC requests massive connections for 5G devices in the range in \cite{ITU2017}, \cite{8476595}. Throughput and latency are the two common KPIs to monitor or differentiate between different services.

On a parallel but different track, the O-RAN Alliance \cite{ORANalliance_2018} defined programmable O-RAN components by dividing functions into smaller modulized elements with advanced virtualization technology and openness \cite{iturria2022multi}. This way, any parties interested in contributing to the RAN industry could interoperate their products and enter the market of intelligent mobile network\cite{Bonati2020}.

\begin{figure}
\centering
  \includegraphics[width=0.8\linewidth]{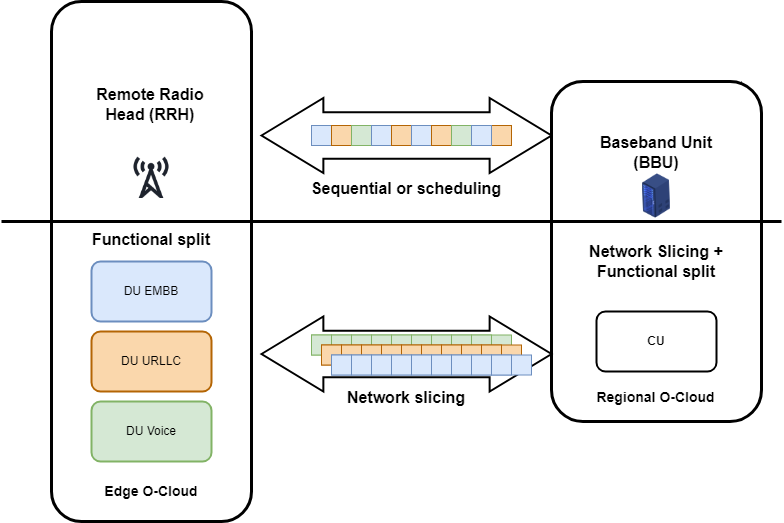}
  \caption{Illustration of functional split and network slicing between RU, DU and CU.}  
\label{fig:functionSplit}
\end{figure}

Functional split between O-RAN components becomes a critical topic \cite{radioaccess2017}. It provides a foundation for O-RAN to design the virtualized disaggregate implementation. For example, the functionality of Baseband Unit (BBU) could be split into a distributed unit (DU) and a centralized unit (CU), where O-RAN takes an extra step of virtualization and turns the components into vDU and vCU. The connection between the DU and CU is called midhaul, or F1-u interface \cite{Hirayama2019}. O-RAN can therefore manage the network slicing within different stages of the functional splits between other modulized components. To handle RAN behaviour between modules and more, RAN Intelligent Controller (RIC) is introduced in O-RAN architecture \cite{RIC6G2021}.

O-RAN empowers resource allocation for network slicing via functional split between different elements in Fig.~\ref{fig:functionSplit}. The traffic with different requirements can have more options to travel along the network. This new functional split concept has shifted the focus of resource allocation solutions from scheduling in frequency and time domain of radio unit (RU) into hardware and software resources toward the BBU side. Thanks to a tremendous amount of work already done on traffic classification, components in O-RAN are able to use known Quality of Service (QoS) labels to allocate the bandwidth. In this research, we focus on allocating the resource specifically on bandwidth sharing on the F1-u interface in disaggregated ORAN architecture.  

Machine Learning (ML) has demonstrated its powerful ability on optimizing the complex 5G wireless network with massive data collected \cite{Elsayed2019}. Therefore, to
 further improve the bandwidth optimization achieved by the functional split, this paper proposes a Reinforcement Learning (RL)-based algorithm, to optimize the target network slice, eMBB, without impacting other slices. The results show it can also improve the QoS of user equipments (UEs) by dynamically changing the bandwidth between network slices on an O-RAN single CU multi DU architecture.

The rest of this paper is organized as follows. In Section II, we review the related works. In Section III, the system model is described in detail, including the integration of functional split and network slicing between O-RAN components. Section IV provides the methodology for fitting the RL model between O-RAN components. Section V shows the result and analysis of our work. Section VI provides the conclusion.

\section{Related Work}
Several ML algorithms are proposed to optimize the physical resource blocks (RBs) between different network slices and use case classes such as eMBB and URLLC. Bektas et al.\cite{Bektas2021} implemented the SAMUS framework to predict incoming traffic. The result provides an insight into different network slicing operations and discussion about the trade-off between the data rate of eMBB or latency of URLLC. Another ML approach solved the localization uncertainty with beam management for the resource allocation on gNB \cite{yao2022deep}. Mollahasani et al. \cite{mollahasani2022energy} implemented a soft actor-critic approach to improve energy efficiency and reduce delay.

Zhang et al. \cite{zhang2022team} proposed a team learning solution for improving the coordination in an O-RAN framework. Thus, their ML algorithm performed a better resource allocation optimization for their access network. Yu et al.\cite{Yu2020} explained different stages of functional splits within 5G RAN. They also proposed a flexible placement for the RAN slicing mapping to reduce the number of active metro nodes. Their result shows that the functional split on the three-layer radio unit (RU)-DU-CU performs better than the two-layer RU-BBU design. Sun et al. \cite{Sun2019} introduced a learning-based model to allocate shared bandwidth between different network slices in a single BBU of C-RAN architecture on the fronthaul interface.

\begin{figure}[h]
 \centering
  \includegraphics[width=0.8\linewidth]{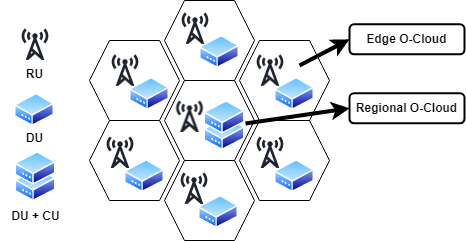}
  \caption{Single CU Multi DU Setup Topology.}  
  \label{fig:honeyComb}
\end{figure}

On top of the mentioned related works, we propose a data-driven solution to make the bandwidth allocation dynamically on the midhaul interfaces between multiple DUs and a single CU shown in Fig. \ref{fig:honeyComb}. Then, we provide an extensive performance evaluation for this approach, proving that our technique yields a better QoS to the moving UEs.

\begin{table}
\centering
\caption{\label{tab:Notations} Notations}
\begin{tabular}{c|c|p{5cm}}
\hline
\textbf{Sets} & \textbf{Size} & \textbf{Description} \\\hline
$i \in \mathcal{I}$ & I &  UEs\\
$j \in \mathcal{J}$ & J &  Edge O-Clouds (DUs)\\
$a \in \mathcal{A}$ & A &  bandwidth change range\\
$k \in \mathcal{K}$ & K &  bandwidth slices\\
$t \in \mathcal{T}$ & T &  time slots\\
$v \in \mathcal{V}$ & V &  UE movement\\
\textbf{Variables} & \textbf{Domain} & \textbf{Description} \\\hline
$b^{t}_{j,k}$ & $[0,100]$ & allocated bandwidth for slice $k$ of DU $j$ \\
$a^{t}_{j,k}$ & $[-10,10]$ & allocated bandwidth percentage change for slice $k$ of DU $j$ \\
\textbf{Given Data} & \textbf{Domain} & \textbf{Description} \\\hline
$v_{i}$ & $\mathbb{N}^{+}$ & movement speed of UE $i$\\
$\beta^{I}_{k}$ & $\mathbb{N}^{+}$ & initial bandwidth for slice $k$\\
$\beta^{M}_{k}$ & $\mathbb{N}^{+}$ & maximum bandwidth for slice $k$\\
$l_{j,k}$ & $\mathbb{N}^{+}$ &  floor value of bandwidth threshold for slice $k$\\
$u^{t}_{j,k}$ & $\mathbb{N}^{+}$ & number of UEs for slice $k$ of DU $j$\\
\textbf{Runtime Data} & \textbf{Domain} & \textbf{Description} \\\hline
$\mathcal{D}^{t}_{i,k}$ & $\mathbb{N}^{+}$ & average data rate of slice $k$ of UE $i$\\
$\mathcal{B}^{t}_{j,k}$ & $\mathbb{N}^{+}$ & throughput of slice $k$ of DU $j$\\
$\mathcal{F}^{t}_{j,k}$ & $\mathbb{N}^{+}$ & Free bandwidth of slice $k$ of DU $j$\\
\end{tabular}
\end{table}

\section{System Model}
Edge and Regional O-Clouds are defined by O-RAN classification group to address different UE requirements in a network \cite{O-RAN-WG62019}. We consider a wireless network following this approach which has $J$ Edge O-Clouds that each one consists of one DU\footnote{Due to having a single dedicated DU in each Edge O-Cloud, we use these two terms interchangeably in this paper.} and one RU\footnote{RU has a radius of 300 meters in 5G, a shorter range compared to older mobile generations but this can vary based on operator settings.}. These Edge O-Clouds connect to a single Regional O-Cloud in the center of these Edge O-Clouds (Fig. \ref{fig:honeyComb}). The Regional O-Cloud includes a Non-Real Time RAN Intelligent Controller (Non-RT RIC) \cite{Bonati2021}, one CU, and an umbrella RU-DU to improve the coverage of the area. This topology design can easily add the UE mobility into the formula where a UE could move within the range of the Edge O-Clouds. Fig.~\ref{fig:functionSplit} illustrates our proposed sliced-based bandwidth optimization between the Edge and Regional O-Cloud. The function split between DU and CU is chosen as the split option 2 defined by 3GPP \cite{radioaccess2017}. 

The group of Edge O-Clouds ($\mathcal{J}$) forms a honeycomb hex in Fig. \ref{fig:honeyComb}. Each Edge O-Cloud ($j\in\mathcal{J}$) connects to the Regional O-Cloud with one midhaul interface. A midhaul interface is divided into $K$ network slices and bandwidth ratio of each slice ($k\in\mathcal{K}$) is represented by $b^{t}_{j,k}$ at time $t$ (Fig.~\ref{fig:functionSplit}). This study defines three ($K=3$) slice types: eMBB, URLLC, and voice. Also, this model may implement mMTC as a fourth slice, but mMTC service type is out of scope in this paper. Each UE ($i\in\mathcal{I}$) belongs to only one of these slice types. Lastly, each UE has a movement type ($v \in \mathcal{V}$), which can be stationary, pedestrian, and vehicle. Table~\ref{tab:Notations} represents the remaining notations used in this paper.

Unlike the other studies on resource allocation using functional split and network slicing \cite{Bektas2021}, \cite{Yu2020}, \cite{Sun2019}, we focus on the throughput optimization problem by applying the RL within a single agent multi nodes model on O-RAN split option 2 between CU and DU on F1-u midhaul interface.

\section{Proposed Solution}

To implement this single sharable agent, the model-free Q-learning is selected as the learning strategy. Q-learning follows the Markov decision process (MDP) by finding an optimal policy from the system described by state ($\mathbb{S}$), action ($\mathbb{A}$), and reward ($\mathbb{R}$). The following are the definition of the three main factors in our system. 

\subsection {Action}
The controller is taking a small step for each Edge O-Cloud $j$ by adding or removing some percentage of the total bandwidth ($a^{t}_{j,k}$) from the network slice $k$ within a given range ($\mathcal{A}$) of total bandwidth at time $t$ represented in Eq.~\eqref{eq:action}. The sum of changes needs to be zero in Eq.~\eqref{eq:sumz} because all network slices share the same midhaul interface in the same Edge O-Cloud shown in Fig.~\ref{fig:functionSplit}. Therefore, the action space in the Q table can be reduced to K-1 slices. Applying this idea to our model, the Q table records actions of eMBB and URLLC slices.  

\begin{flalign}
&\mathbb{A}(t)= \{ \{a^{t}_{j,k} \in\mathcal{A}  | j\in\mathcal{J}, k\in\mathcal{K}^{-} \} \label{eq:action}\\
&\sum\limits_{k=0}^{K} a^{t}_{j,k}= 0,\quad \forall j\in\mathcal{J}, \forall t\in\mathcal{T}\label{eq:sumz}
\end{flalign}

\subsection {State}
In our proposed scheme, in the Q table, the state can be represented by two components in the Edge O-Cloud \(j\) in Eq.~\eqref{eq:state}. 
\begin{itemize}
\item Percentage of total bandwidth allocation \(b_{j,k}\) for each network slice \(k\) on the shared midhaul interface.
\item User counts per slice \(u_{j,k}\).
\end{itemize} 

The state $\mathbb{S}_{j}(t)$ at timestamp $t$ is described by the bandwidth changed from earlier action $\mathbb{A}(t-1)$. In other words, the timestamp of $t-1$ here represents the state at the moment before the action $\mathbb{A}(t)$ is applied to the bandwidth. 
\begin{flalign}
    &\mathbb{S}_{j}(t)= \{ \{b^{(t-1)}_{j,k} | k\in\mathcal{K}\},
                                    \{u^{(t-1)}_{j,k}| k\in\mathcal{K} \} \}
    \label{eq:state}
\end{flalign}

\subsection {Reward}
Increasing the eMBB throughput is one of the main goals of this study. Therefore, the controller calculates the reward based heavily on the Edge O-Cloud throughput of the eMBB rate and the average eMBB QoS on UE. The negative part of the reward on Eq. \eqref{eq:panalty} is to guarantee a minimum bandwidth ($L$) for the network slices. Therefore, the network slice will always have some free bandwidth for the subsequent UE request. The positive part of the reward on Eq. \eqref{eq:reward} is the sum of the following three components, the increase of free bandwidth ($\mathcal{F}$), current used bandwidth ($\mathcal{B}$), and UE data rate ($\mathcal{D}$). Therefore, the reward combines the negative and positive portion in Eq. \eqref{eq:totalR}. The scale factor($s$) and weight factor ($w$) are both implemented to maintain the component calculated on the same scale and rank each component's importance.

\begin{flalign}
& 
\mathbb{N}_{j}(t)=\sum\limits_{k \in\mathcal{K}} \begin{cases}
 1 & \text{if }b^{t}_{j,k} < L \\
 0 & \text{otherwise}
\end{cases} \label{eq:panalty}\\
& \mathbb{P}_{j}(t)= \sum\limits_{k \in\mathcal{K} }(s_{1,k}\Delta\mathcal{F}^{t}_{j,k}+s_{2,k}\mathcal{B}^{t}_{j,k}+s_{3,k}\mathcal{D}^{t}_{j,k}) \label{eq:reward}\\
& \mathbb{R}_{j}(t)=w_{p}\mathbb{P}_{j}(t)-w_{n}\mathbb{N}_{j}(t) \label{eq:totalR}
\end{flalign}

\begin{algorithm}[!hbt]
 \caption{Q-learning Based Solution}
 \label{alg:the_alg}
 \begin{algorithmic}[1]
 \renewcommand{\algorithmicforall}{\textbf{For Each}}

 \STATE Initialization: 
 \STATE Each DU $j$ starts with bandwidth $b_{j,k}$ for each $k$ slice
 \STATE Each UE $i$ is randomly assigned with moving speed $v_i$ m/s and network slice $k$
 \STATE Each UE $i$ connects to nearby DU $j$ and requesting $y_i$ GB data
  \FORALL{environment step at time $t$}
    \STATE UE ($i$) moves $v_i$ m/s, receives $d_i$ G/s
    \STATE Controller observes state $s_t$
    \STATE x $\leftarrow$ uniform random number between 0 and 1
    \IF{ $x \geq \epsilon$ }
        \STATE Select $A^t_j = \biggl[ a^t_{k1,j}, a^t_{k2,j}... \biggr] =\arg\max_{a} Q_(s_t, a_t)$ for each DU $j$
    \ELSE
        \STATE Select random action $A^t_j$ for each DU $j$
    \ENDIF
    \STATE Execute $A^t_j$ for each DU $j$
    \STATE Observe the next state $s_{t+1}$ and reward $r_{t+1}$
    \STATE Update Value Action: $Q(s_{t},a_{t}) \leftarrow (1-\alpha) Q(s_{t},a_{t})$ 
    \STATE $+ \alpha \biggl [ r_{t+1}+\gamma \max\limits_{A} Q(s_{t+1},A)\biggr ]$

  \ENDFOR
 \end{algorithmic} 
 \end{algorithm}
 
 \subsection{Sharable Q Table Approaches}
All DUs share the same Q table by referencing the table from its states and actions. In the Q table, the state space is composed by the bandwidth in percentage \(b_{j,k}\) of each network slice $k$ and the connected user counts \(u_{j,k}\) on Edge O-Cloud $j$. This state representation allows the Q table to be shared and updated by all $J$ DUs independently in each step. This sharing strategy is based on similar environment settings in the neighboring DUs. The important variable is the incoming UE requests which will impact the required bandwidth of different network slices accordingly. Hence, the bandwidth distribution of slices and user counts are selected as the state of the Q table. These parameters change over time in each DU during the simulation by UEs' movement and QoS types. Thus, each DU creates a specific scenario or state that could apply to neighbored DUs later. Additionally, the RL controller can rapidly fill the Q table by randomly selecting actions for each DU. 

As described in Algorithm~\ref{alg:the_alg}, the Q learning model is trained for $N$ episodes, and each episode contains $T$ seconds. On initialization, each network slice $k$ of each DU $j$ starts with $\beta_{j,k}$ percent of bandwidth. Each UE $i$ is assigned to a fixed data amount $y_i$, a network slice $k$, and a movement type defined in Table~\ref{table:envSetup}. After the simulation begins, several events happen simultaneously on each second $(t)$.
\begin{itemize}
\item Each UE $i$ moves $v_i$ meter based on its movement type, which is independent of the QoS traffic type. The UE $i$ also receives $d_i$ GB data after connecting to a nearby Edge O-Cloud $j$. The sum of connected UEs' data rate $\sum d_i$ will be the used as the throughput $\mathcal{B}$ of the DU $j$.
\item On the Regional O-Cloud, the Non-RT RIC will check the incoming request of traffic for each DU $j$ as a state $\mathbb{S}_{j}(t)$. Based on an epsilon-greedy approach ($\epsilon$), the RIC will randomly pick an action $a_{j,k}\in{A_j}$ or look up from the Q table to find the best action for the Edge O-Cloud to take. The action $a_{j,k}$ causes the increasing or decreasing the bandwidth $b_{j,k}$ of a network slice $k$. After the action is taken, the Edge O-Cloud will update the Q table accordingly with reward $\mathbb{R}_j(t)$.  
\end{itemize}

\begin{table}[h]
\centering
\caption{\label{table:envSetup} Environment Parameters}
\begin{tabular}{l|lll}
\hline
\textbf{Variables}      & \multicolumn{3}{l}{\textbf{Unit}}               \\ \hline
eMBB Initial Rate       & \multicolumn{3}{l}{1 GB/s}                      \\
URLLC Inital Rate       & \multicolumn{3}{l}{100 MB/s}                    \\
Voice Initial Rate      & \multicolumn{3}{l}{100 MB/s}                    \\
eMBB Max Rate           & \multicolumn{3}{l}{N/A}                         \\
URLLC Max Rat           & \multicolumn{3}{l}{5 GB/s}                      \\
Voice Max Rate          & \multicolumn{3}{l}{1 GB/s}                      \\ \hline
\textbf{Movement Speed} & \multicolumn{3}{l}{}                            \\ \hline
Stationary Speed        & \multicolumn{3}{l}{0 km/hr}                     \\
Pedestrian Speed        & \multicolumn{3}{l}{0 km/hr to 10 km/hr}         \\
Vehicular Speed         & \multicolumn{3}{l}{10 km/hr to 120 km/hr}       \\ \hline
\textbf{Testset}        & \textbf{Mid}  & \textbf{High}  &                \\ \hline
eMBB UEs                & \multicolumn{3}{l}{{[}100,200,300,400,500{]}}   \\
URLLC UEs               & 500           & 1000           &                \\
Voice UEs               & 1000          & 500            &                \\ \hline
\textbf{Bandwidth (\%)} & \textbf{eMBB} & \textbf{URLLC} & \textbf{Voice} \\ \hline
Balanced                & 40            & 30             & 30             \\
eMBB Focus              & 90            & 5              & 5
\end{tabular}
\end{table}

\begin{figure}[hbt!]
 \centering
  \includegraphics[width=0.5\textwidth]{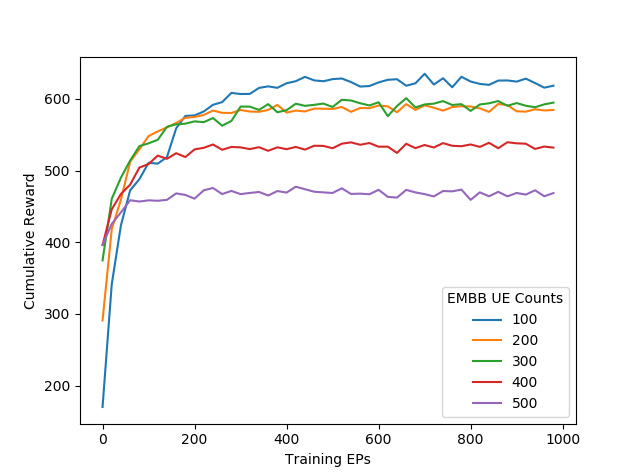}
  \caption{Reward convergence during training process.}  
  \label{fig:rewardConv}
\end{figure}

\section{Performance Evaluation}

\subsection{Simulation Platform} 
The simulation environment is implemented with a Python framework, SimPy\cite{simpy}. The framework is designed for handling "process-based discrete-event," which is ideal for our heavy need of scheduling and queuing in the edge network to handle various UE events and data transmission.
The RL environment is implemented in OpenAI gym environment \cite{openai}. The gym environment is able to stack and interact with the SimPy network environment and can be easily applied with other RL algorithms in the future. The implemented network slices are eMBB, URLLC and Voice.

The environment has the following parameters defined in Table~\ref{table:envSetup}. Each network slice has its initial data rate to assign to newly connected UEs. Meanwhile, these UEs may demand more data rate if connected Edge O-Cloud has free bandwidth in the dedicated slice. Each UE movement type has a given range of movement speed per second according to \cite{ITU2017}. By taking action every second with the same time unit used by UE movement, the reward can be calculated for evaluation on every second. To evaluate the performance of our model in different traffic demands, two sets of UE parameters are proposed. In the first case, URLLC has 500 UEs, and voice has 1000 UEs, which is called the "Mid traffic profile." In the second case, "High traffic profile," URLLC has 1000 UEs, and voice has 500 UEs. The naming of profiles is based on the traffic demand differently between these two slice types. By the definition of those services, URLLC is known to request higher data than Voice data; thus, increasing the number of URLLC UEs will lead to more traffic demand to the network. In both cases, the number of eMBB UEs increases from 100 UEs to 500 UEs. With concept of stress test, increasing 100 eMBB UEs each time can show the model's response to the significant increase in traffic demand.

\begin{figure*}[]
\centering
\begin{minipage}{.8\columnwidth}
  \centering
  \includegraphics[width=\textwidth]{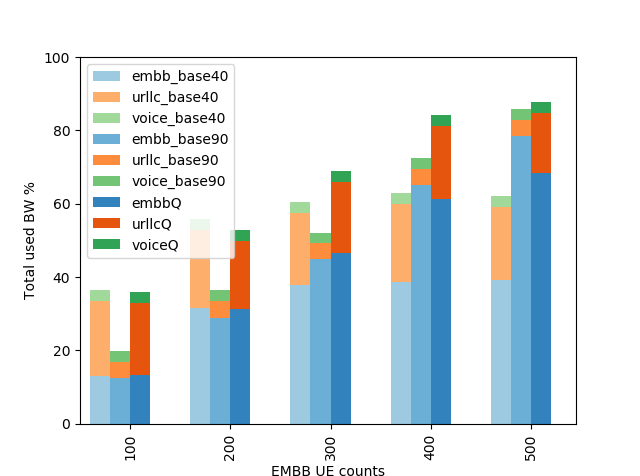}
  \caption*{(a) Comparison in Mid URLLC.}  
\end{minipage}%
\begin{minipage}{.8\columnwidth}
  \centering
  \includegraphics[width=\textwidth]{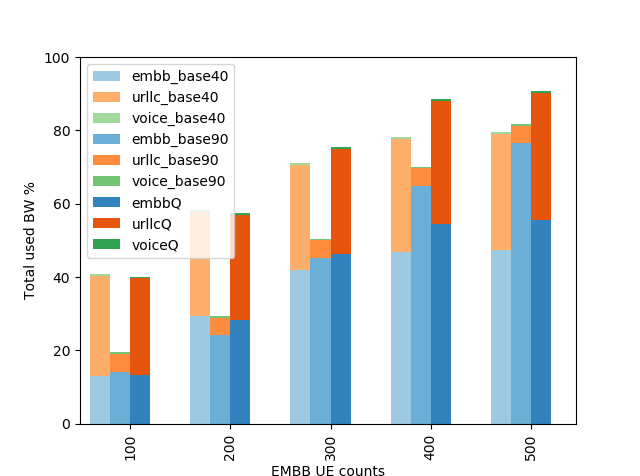}
  \caption*{(b) Comparison in High URLLC.}  
\end{minipage}%
\caption{Bandwidth distribution in BBU.}
\label{fig:BBUresult}
\end{figure*}

In addition, two sets of baselines (Table~\ref{table:envSetup}) are proposed to evaluate the proposed solution. The first one is the "Balanced" profile, which has a slightly higher bandwidth distribution to eMBB of 40\% while URLLC and Voice have 30\% of bandwidth allocated. The other one is the "eMBB Focus" profile. This profile gives 90\% bandwidth to the eMBB slice.

The proposed Q-learning model is trained on different eMBB UEs defined in Table~\ref{table:envSetup}. Fig. \ref{fig:rewardConv} shows the learning curves of these scenarios. Each use case can converge to the optimum solution after around 400 iterations.

\subsection{Results and Analysis}
Fig.~\ref{fig:BBUresult} illustrates the comparison between two baselines and our Q-learning model on 120 second simulation on O-Clouds. For each eMBB user scenario, three bars represent the bandwidth allocating result of Balanced, eMBB focus, and Q-learning from left to right. In Fig.~\ref{fig:BBUresult}, both mid or high traffic profiles show the same trend. The figure illustrates that the balanced baseline can provide sufficient bandwidth to all network slices up to 300 eMBB UEs. This balanced-fixed bandwidth approach is inadequate for a higher number of eMBB users starting from 400 eMBB UEs. On the other hand, the eMBB focus profile, base90, cannot provide enough bandwidth for URLLC slice since the beginning of 100 eMBB UEs, but it has the highest eMBB throughput on 500 eMBB UEs. In other words, at least one network slice may not be able to demand more bandwidth from the Edge O-Cloud on one of the baselines. Conversely, the Q-learning model can dynamically assign unused portions from less busy slices to high traffic demand slice(s). Fig.~\ref{fig:BBUresult} shows eMBB always has the highest bandwidth assigned while URLLC and the voice have adequate bandwidth allocated. In addition, the total bandwidth utilization is higher on Q-learning compared to both balanced and eMBB focus baselines by at least 5\%. The Q-learning model provides a better bandwidth distribution with fairness and total bandwidth utilization to handle all network slices on the throughput comparison. 

\begin{figure*}[htb!]
\centering
\begin{minipage}{.8\columnwidth}
  \centering
  \includegraphics[width=\textwidth]{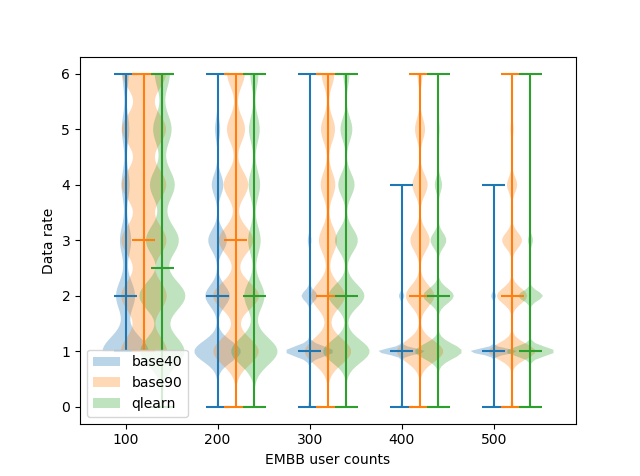}
  \caption*{(a) Comparison in Mid URLLC.}  
\end{minipage}%
\begin{minipage}{.8\columnwidth}
  \centering
  \includegraphics[width=\textwidth]{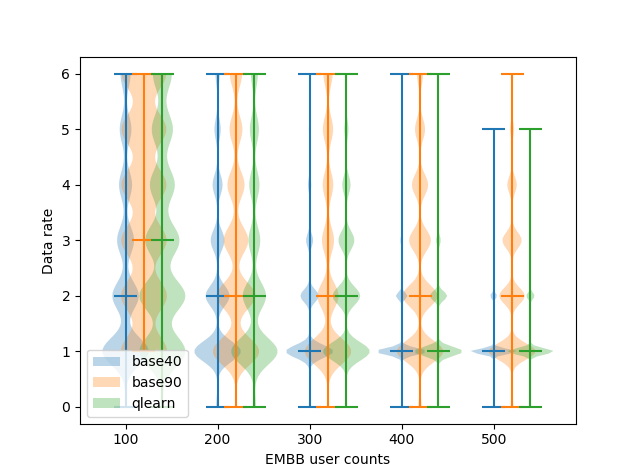}
  \caption*{(b) Comparison in High URLLC.}  
\end{minipage}%
\caption{UE eMBB data rate distribution.}
\label{fig:eMBBUEDR}
\end{figure*}

\begin{figure*}[htb!]
\centering
\begin{minipage}{.8\columnwidth}
  \centering
  \includegraphics[width=\textwidth]{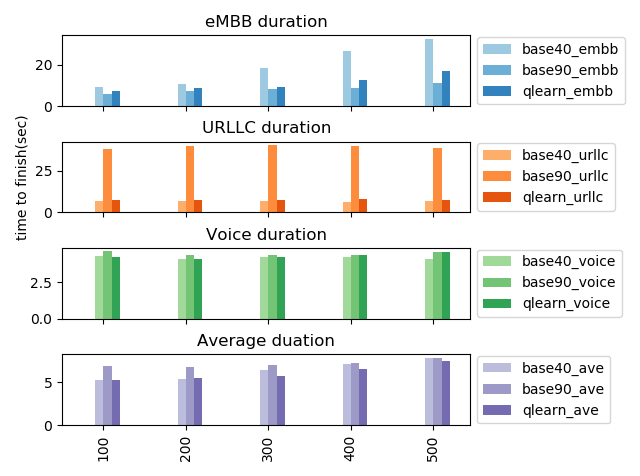}
  \caption*{(a) Comparison in Mid URLLC.}  
\end{minipage}%
\begin{minipage}{.8\columnwidth}
  \centering
  \includegraphics[width=\textwidth]{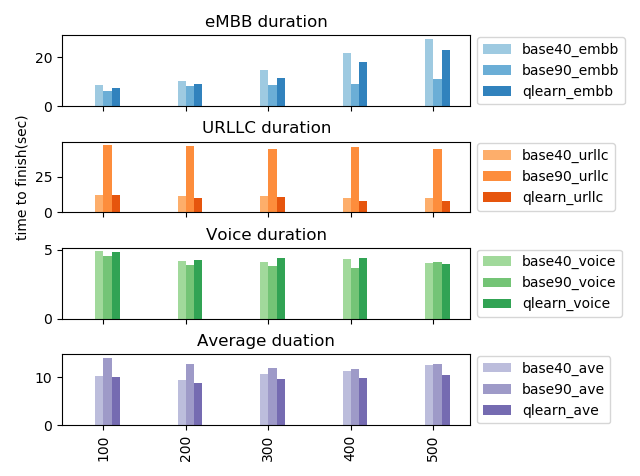}
  \caption*{(b) Comparison in High URLLC.}  
\end{minipage}%
\caption{Time to Finish on UE.}
\label{fig:TTF}
\end{figure*}

Meanwhile, the simulation also reflects the QoS of UEs. Data rate and time to finish static data requests are two good QoS indicators. We illustrate the result of UE experiences in Fig. \ref{fig:eMBBUEDR} and Fig. \ref{fig:TTF}. 

Fig.~\ref{fig:eMBBUEDR} shows the eMBB user data rate distribution during the simulation. The result of the balanced baseline in blue illustrates a degradation of the eMBB peak rate when the eMBB user count increases in the network. The eMBB focus baseline in orange shows its ability to provide the best QoS for eMBB users without impacting the peak rate while eMBB users increase in the network. Q-learning model in green can uphold the peak eMBB rate regardlessly in Fig.~\ref{fig:eMBBUEDR}(a). This shows that minor loss of bandwidth is not impacting the QoS much in the scope of the peak rate. However, Q-learning is not able to provide a good peak rate when reaching the 500 eMBB users in the high traffic scenario compared to the eMBB focus baseline in Fig.~\ref{fig:eMBBUEDR}(b). We propose that the degradation of the peak rate is acceptable here compared to the saturated bandwidth of URLLC on the Edge O-Cloud side.

In the result of transmission time in Fig. \ref{fig:TTF}, the balanced baseline has a similar transmission duration compared to the result of Q-learning. On the other hand, the eMBB focus baseline takes a significantly longer time to finish for URLLC data which is against the definition of URLLC. On average duration, Q-learning can finish the transmission faster than baseline in both balanced and eMBB focus baselines. This indicates that the minor throughput loss on eMBB in Q-learning model compared to the eMBB focus baseline on the Edge O-Cloud side is not negatively impacting the QoS of UE, especially not on the transmission time. This means the Q-learning solution gives a better bandwidth fairness solution than the eMBB focus scenario when Edge O-Cloud has high traffic demand from all slices. Compared to the balanced scenario, the Q-learning model can provide a better eMBB QoS solution by increasing the throughput of eMBB as much as possible without impacting throughput and QoS on other slices.

\section{Conclusion}

This study proposes an RL solution to optimize resource allocation by integrating O-RAN functional split and network slicing. The contribution of our work focuses on bandwidth optimization on a 5G O-RAN single CU multi DU architecture. With a sharable Q table in Q-learning, our model gives a data-driven solution for bandwidth optimization on disaggregated 5G O-RAN with fairness to the throughput of network slices and improves the QoS of UEs. In our future work, we plan to use other RL techniques and extend the design in this paper to improve QoS for high-speed vehicles.

\bibliography{main,main2}

\begin{thebibliography}{10}
\providecommand{\url}[1]{#1}
\csname url@samestyle\endcsname
\providecommand{\newblock}{\relax}
\providecommand{\bibinfo}[2]{#2}
\providecommand{\BIBentrySTDinterwordspacing}{\spaceskip=0pt\relax}
\providecommand{\BIBentryALTinterwordstretchfactor}{4}
\providecommand{\BIBentryALTinterwordspacing}{\spaceskip=\fontdimen2\font plus
\BIBentryALTinterwordstretchfactor\fontdimen3\font minus
  \fontdimen4\font\relax}
\providecommand{\BIBforeignlanguage}[2]{{%
\expandafter\ifx\csname l@#1\endcsname\relax
\typeout{** WARNING: IEEEtran.bst: No hyphenation pattern has been}%
\typeout{** loaded for the language `#1'. Using the pattern for}%
\typeout{** the default language instead.}%
\else
\language=\csname l@#1\endcsname
\fi
#2}}
\providecommand{\BIBdecl}{\relax}
\BIBdecl

\bibitem{3GPPREl16}
``Summary of release 16 work items,'' 3GPP, TR 21.916, 2021.

\bibitem{ITU2017}
``Minimum requirements related to technical performance for imt-2020 radio
  interface(s),'' ITU-R, TR IMT-2020.TECH PERF REQ, Nov 2017.

\bibitem{8476595}
P.~Popovski, K.~F. Trillingsgaard, O.~Simeone, and G.~Durisi, ``5g wireless
  network slicing for embb, urllc, and mmtc: A communication-theoretic view,''
  \emph{IEEE Access}, vol.~6, pp. 55\,765--55\,779, 2018.

\bibitem{ORANalliance_2018}
\BIBentryALTinterwordspacing
``O-ran alliance: About us,'' O-RAN Alliance, Feb 2018. [Online]. Available:
  \url{https://www.o-ran.org/about .}
\BIBentrySTDinterwordspacing

\bibitem{iturria2022multi}
P.~E. Iturria-Rivera, H.~Zhang, H.~Zhou, S.~Mollahasani, and M.~Erol-Kantarci,
  ``Multi-agent team learning in virtualized open radio access networks
  (o-ran),'' \emph{Sensors}, vol.~22, no.~14, p. 5375, 2022.

\bibitem{Bonati2020}
L.~Bonati, M.~Polese, S.~D’Oro, S.~Basagni, and T.~Melodia, ``Open,
  programmable, and virtualized 5g networks: State-of-the-art and the road
  ahead,'' \emph{Computer Networks}, vol. 182, p. 107516, 12 2020.

\bibitem{radioaccess2017}
A.~Umesh, ``Study on new radio access technology: Radio access architecture and
  interfaces (release 14),'' 3GPP, TR 38.801, Apr 2017.

\bibitem{Hirayama2019}
H.~Hirayama, Y.~Tsukamoto, S.~Nanba, and K.~Nishimura, ``{RAN Slicing in
  Multi-CU/DU Architecture for 5G Services},'' in \emph{2019 IEEE 90th
  Vehicular Technology Conference (VTC2019-Fall)}, vol. 2019-Septe.\hskip 1em
  plus 0.5em minus 0.4em\relax IEEE, sep 2019, pp. 1--5.

\bibitem{RIC6G2021}
M.~Dryjański, Łukasz Kułacz, and A.~Kliks, ``Toward modular and flexible
  open ran implementations in 6g networks: Traffic steering use case and o-ran
  xapps,'' \emph{Sensors}, vol.~21, p. 8173, 12 2021.

\bibitem{Elsayed2019}
M.~Elsayed and M.~Erol-Kantarci, ``Ai-enabled future wireless networks:
  Challenges, opportunities, and open issues,'' \emph{IEEE Vehicular Technology
  Magazine}, vol.~14, pp. 70--77, 9 2019.

\bibitem{Bektas2021}
C.~Bektas, D.~Overbeck, and C.~Wietfeld, ``{SAMUS: Slice-Aware Machine
  Learning-based Ultra-Reliable Scheduling},'' in \emph{ICC 2021 - IEEE
  International Conference on Communications}.\hskip 1em plus 0.5em minus
  0.4em\relax IEEE, jun 2021, pp. 1--6.

\bibitem{yao2022deep}
Y.~Yao, H.~Zhou, and M.~Erol-Kantarci, ``Deep reinforcement learning-based
  radio resource allocation and beam management under location uncertainty in
  5g mmwave networks,'' in \emph{IEEE Symposium on Computers and Communications
  (ISCC)}, 2022.

\bibitem{mollahasani2022energy}
S.~Mollahasani, T.~Pamuklu, R.~Wilson, and M.~Erol-Kantarci, ``Energy-aware
  dynamic du selection and nf relocation in o-ran using actor--critic
  learning,'' \emph{Sensors}, vol.~22, no.~13, p. 5029, 2022.

\bibitem{zhang2022team}
H.~Zhang, H.~Zhou, and M.~Erol-Kantarci, ``Team learning-based resource
  allocation for open radio access network (o-ran),'' in \emph{IEEE
  International Conference on Communications (ICC)}, 2022.

\bibitem{Yu2020}
H.~Yu, F.~Musumeci, J.~Zhang, M.~Tornatore, and Y.~Ji, ``{Isolation-Aware 5G
  RAN Slice Mapping over WDM Metro-Aggregation Networks},'' \emph{Journal of
  Lightwave Technology}, vol.~38, no.~6, pp. 1125--1137, 2020.

\bibitem{Sun2019}
Y.~Sun, Y.~Wang, H.~Yu, B.~Guo, and X.~Zhang, ``{A learning-based bandwidth
  resource allocation method in sliced 5G C-RAN},'' \emph{2019 IEEE Globecom
  Workshops, GC Wkshps 2019 - Proceedings}, 2019.

\bibitem{O-RAN-WG62019}
O-RAN-WG6, ``{Cloud architecture and deployment scenarios for O-RAN virtualized
  RAN - v02.02},'' Tech. Rep., 2021.

\bibitem{Bonati2021}
L.~Bonati, S.~D'Oro, M.~Polese, S.~Basagni, and T.~Melodia, ``Intelligence and
  learning in o-ran for data-driven nextg cellular networks,'' \emph{IEEE
  Communications Magazine}, vol.~59, pp. 21--27, 10 2021.

\bibitem{simpy}
\BIBentryALTinterwordspacing
``Discrete event simulation for python,'' SimPy, 2020. [Online]. Available:
  \url{https://simpy.readthedocs.io/en/latest/}
\BIBentrySTDinterwordspacing

\bibitem{openai}
\BIBentryALTinterwordspacing
``Build next-gen apps with openai’s powerful models,'' OpenAI, 2022.
  [Online]. Available: \url{https://openai.com/api/}
\BIBentrySTDinterwordspacing

\end{thebibliography}

\end{document}